\documentstyle[twocolumn,aps]{revtex}

\def\ga{\mathrel{\vcenter{\offinterlineskip\halign{\hfil
$\displaystyle##$\hfill\cr>\cr\sim\cr}}}}

\def\omegabar{\hbox{${\bar{\omega}}$}}
\def\boldom{\hbox{\pmb{$\omega$}}}
\def\boldthet{\hbox{\pmb{$\theta$}}}
\def\boldphi{\hbox{\pmb{$\phi$}}}

\def\pmb#1{\setbox0=\hbox{#1}
   \kern-0.025em\copy0\kern-\wd0
   \kern.05em\copy0\kern-\wd0
   \kern-.025em\raise.0433em\box0}

\begin{document}

\draft
\preprint{UF-IFT-HEP-97-11;~astro-ph/9707196\\}

\title{Relativistic electrons in a
rotating spherical magnetic dipole:~localized three-dimensional states\\
}

\author{James M. Gelb\cite{gelb}\\}
\address{
Department of Physics\\
University of Texas at Arlington\\
Arlington, Texas 76019\\
}
\author{Kaundinya S. Gopinath and Dallas C. Kennedy\cite{kennedy}\\}
\address{
Department of Physics\\
University of Florida at Gainesville\\
Gainesville, Florida 32611\\
}

\date{\today}
\maketitle

\begin{abstract}
Paralleling a previous paper, we examine single- and many-body states of
relativistic electrons in an intense, rotating magnetic dipole field.  
Single-body orbitals are derived semiclassically and then
applied to the many-body case via the Thomas-Fermi approximation.  The 
many-body case is reminiscent of the quantum Hall state.
Electrons in a realistic neutron star crust are considered with both
fixed density profiles and constant Fermi energy.  In the first case, 
applicable to young neutron star crusts, the
varying magnetic field and relativistic Coriolis correction lead to a varying 
Fermi energy and macroscopic currents.  
In the second, relevant to older crusts, the electron density is redistributed
by the magnetic field.
\end{abstract}


\narrowtext

\section*{1. INTRODUCTION}

In a previous paper~\cite{gelb98} (hereafter paper I),
we examined the relativistic semiclassical relativistic orbitals of a particle
of mass $m$ and charge $q$ in an intense magnetic field, idealized as a 
dipole, in a rotating reference frame.  The particle was 
confined to the spherical surface.  In this paper, we present 
the treatment of three-dimensional orbitals, 
the local cyclotron or Landau states.  These results are applied to 
many-body electron states defined in a semiclassical (Thomas-Fermi) 
approximation and calculated in a simplified neutron star crust model with 
electrons and nuclei.  We consider quantum dynamics only for the electrons,
not the hadrons~\cite{chak97}, but effects of
relativity and rotation are included.

We ignore gravity (derivatives of the metric) here, as this is negligible for
charged particles compared to the magnetic field.  Where numerical
values are needed, the tilt angle $\theta_0$ between the dipole and rotation 
axes is assumed to be maximal, $\sin\theta_0$ = 1, and the rotational velocity
$\Omega$ to be $\omegabar$ = $\Omega R/c$ = 0.01, a realistic value
for high-field neutron stars with radius $R$~\cite{glen97,shap83}.

The treatment is based on expanding the 
particle motion in inverse powers of the field strength.  Although electrons
are stripped from the neutron star surface by the rotation-induced electric
field, the bulk of electrons remain in the crust to preserve local charge
neutrality, with the surface sheathed by a thin space charge.  The space charge
is stabilized by the Coulomb force (with the positive crystal) opposing the
induced electric field.  Only a small fraction of electrons 
are accelerated into the stellar wind~\cite{mich91}. Our final 
applications here are to quantum single- and many-body states
where radiation emission is neglected.  This is exact for charged particles
in their ground states or in excited states unable to decay by Pauli exclusion
blocking in the presence of other fermions.
We also seek a general classification of possible orbitals based 
on the relevant kinematic parameters.  The semiclassical quantization is
based on the Wilson-Sommerfeld or Bohr-Sommerfeld rule, a result of
the WKB approximation~\cite{land77}.

\begin{figure}
\vspace{11pc}
\caption{
Geometry of the magnetic dipole $\bf{M}$ and sphere rotating at angular 
velocity $\bf{\Omega}$, with relative tilt $\theta_0$.}
\label{fig1}
\end{figure}

\section*{2. GENERAL RELATIONS}

\subsection*{2.1 Coordinates, metric, and field}

As the magnetic field dominates rotational effects (unless $r$ is quite large),
we place the rotation axis at an angle $\theta_0$ with respect to the magnetic
dipole in the $\phi$ = 0, $\pi$ plane (Fig.~1).  The metric in a spherical 
polar coordinate system $(r,\theta,\phi )$ rotating with the sphere is 
given by the line element
\begin{eqnarray}
ds^{2} &=& g_{\mu\nu}dx^{\mu}dx^{\nu}\nonumber\\
 &=&c^{2}(1-\boldom^2)dt^{2}-dr^2
-r^{2}(d\theta^{2}+\sin^{2}\theta d\phi^{2}) - \nonumber\\
 & &2cr\omega_\phi\sin\theta~dt~d\phi - 2cr\omega_\theta~dt~d\theta \quad .
\end{eqnarray}
The vector $\boldom$ is defined from the rotational angular velocity vector
$\mathbf{\Omega}$ by $\boldom$ = $\mbox{\boldmath $\Omega\times r$}/c$.  We
also use $\omegabar\equiv$ $\Omega R/c,$ where $r$ = $R$ is a reference sphere.
The components of $\boldom$ are
\begin{eqnarray}
\omega_{\phi} & = & (\omegabar r/R)[\cos\theta_0\sin\theta - 
\sin\theta_0\cos\theta\cos\phi ]\quad , \nonumber \\
\omega_{\theta} & = & -(\omegabar r/R)\sin\theta_0\sin\phi\quad , \\
{\boldom}^2 & = & \omega^2_\phi + \omega^2_\theta\quad . \nonumber
\end{eqnarray}
These are the appropriate generalizations of paper~I and the 
treatment of Landau \& Lifshitz\cite{land75}
to the case $\sin\theta_0\neq$ 0 and $r\neq$ constant.  The lightsphere is the
surface $\boldom^2$ = 1, or
\begin{eqnarray}
\lefteqn{\omegabar^2(r/R)^2[\cos^2\theta_0\sin^2\theta +}\nonumber\\
 & & \sin^2\theta_0\sin^2\phi \sin^2\theta_0\cos^2\theta\cos^2\phi -\nonumber\\
 & & (1/2)\sin 2\theta_0\sin 2\theta\cos\phi ] = 1\quad .
\end{eqnarray}
This is the surface upon which $g_{00}$ = 0 (Fig.~2).

We choose our axes so that the magnetic dipole is along the $\theta = 0$ 
direction.  The dipole magnetic field has polar strength $B_0$ at
$r$ = $R:$
\begin{eqnarray}
A_{\theta} & = & 0 \\
A_{r} & = & 0 \\
A_{\phi} & = & \frac{B_{0}R^3}{2r}\sin^2 \theta\quad ,
\end{eqnarray}
\noindent
where $A_\phi$ is covariant in the rotating spherical coordinates.  For
convenience, we rescale $B_0$ into dimensionless form as 
\begin{eqnarray}
\beta_0 = |q|B_0R/(2mc^2)\quad .
\end{eqnarray}
The magnetic moment of this dipole field is
\begin{equation}
|{\bf M}| = B_0 R^3/2 \quad .
\end{equation}
In a high-field neutron star, $\beta_0\sim$ $10^{15}.$

The Lagrangian of the charged particle is expressed in terms of the proper 
time $\tau ,$ 
with $\dot{x^\mu}\equiv$ $dx^\mu /d\tau .$  There are four equations of
motion including one for each momentum component and one energy-momentum 
constraint.
Since the Lagrangian does not explicitly depend on $t,$ we have
\begin{equation}
\frac{\partial L}{\partial t}=0\quad .   
\end{equation}
\noindent
which implies
\begin{equation}
\frac{dP_{0}}{d\tau}=0\quad .
\end{equation}
\noindent
Because
\begin{equation}
d\tau=\frac{dt\sqrt{g_{\mu\nu}(dx^{\mu}/dt)(dx^{\nu}/dt)}}{c}\quad ,
\end{equation}
\noindent
we have also
\begin{equation}
\frac{dP_{0}}{dt}=0\quad ,
\end{equation}
\noindent where $P_0$ = $E,$ the energy.
The equations of motion for $r,$ $\phi ,$ and $\theta$ are non-trivial:
\begin{equation}
\frac{\partial L}{\partial r}-\frac{dP_{r}}{d\tau}\ =\
\frac{\partial L}{\partial \phi}-\frac{dP_{\phi}}{d\tau}\ =\
\frac{\partial L}{\partial \theta}-\frac{dP_{\theta}}{d\tau}\ =\ 0\quad .
\end{equation}
\noindent

The energy-momentum-mass constraint is
\begin{eqnarray}
g^{00}P_{0}^{2} + 2g^{0i}P_{0}(P_{i}-\frac{q}{c}A_{i}) +g^{ij}(P_{i}-
\frac{q}{c}A_{i})(P_{j}-\frac{q}{c}A_{j})\nonumber\\
 = (mc)^{2}~, &
\end{eqnarray}
\noindent
including radial and angular terms, and the mixed rotational 
$g^{0\theta},$ $g^{0\phi}$ terms.  The contravariant metric components are
\begin{eqnarray}
g^{00} = 1/c^2~, & g^{rr} = -1/r^2~, \nonumber \\
g^{\theta\theta} = -(1-\omega_{\theta}^2)/r^2~, &
\ \ g^{\phi\phi} = -(1-\omega_{\phi}^2)/(r^2\sin^2\theta)~, \\
g^{0\phi} = -\omega_\phi/(cr\sin\theta)~, &
g^{0\theta} = -\omega_\theta/(cr)~. \nonumber
\end{eqnarray}
\noindent
Along the actual worldpath in spacetime the energy-momentum constraint is
equivalent to $g_{\mu\nu}\dot{x^\mu}\dot{x^\nu}$ = $c^2.$  This condition is 
valid after varying the action and simplifies the equations of motion.

\subsection*{2.2 Asymptotic orbits}

As a check of this dynamical system, we consider briefly the $r\rightarrow$
$\infty$ orbits.  The constraint~(14) alone is sufficient to give
essential information about the trajectory.  Note first that $\boldom^2<$ 1
for physical trajectories.  Since $\boldom^2$ = $\omega^2_\phi$ + 
$\omega^2_\theta$, each component must also have magnitude less than one.
Then $g^{\theta\theta}$ and $g^{\phi\phi}$ are never exactly zero.

In paper~I the natural energy scale for unconfined charged particles at the
surface $r$ = $R$ was set by $\epsilon$ $\sim\beta_0,$ or $E$ $\sim |q|B_0R/2,$
independent of $m$ (see also Longair~\cite{long92}).  In the limit 
$r\rightarrow$ $\infty$, Eq.~(14) simplifies to
\begin{eqnarray}
P^2_r = (E/c)^2 - (mc)^2\quad ,
\end{eqnarray}
where $E$ is naturally $\ga{\cal O}(|q|B_0R).$  In this regime, $B_0$ is
much smaller than at the surface of the star, as $r$ is much larger.
The magnetic potential term vanishes as $1/r^2,$ while the angular momentum
terms vanish as $1/r.$  Particles can escape as $r\rightarrow$ $\infty$ 
if $P^2_r >$ 0 and if they remain within the lightsphere surface $\boldom^2$ 
= 1 (Fig.~2).
Both graphically and analytically, it is seen that $(\phi ,\theta )\rightarrow$
$(0,\theta_0)$ or $(\pi ,\pi -\theta_0),$ as $r\rightarrow$ $\infty$.
That is, an escaping particle is forced into the common plane of ${\bf M}$ and
${\bf \Omega}$ and leaves along the rotation axis.

\begin{figure}
\vspace{11pc}
\caption{
Lightsphere $r(\theta ,\phi )$ defined by~Eq.(3), surface of 
$g_{00}$ = 0, for $\omegabar$ = 0.01.  $\theta$ and $\phi$ in radians.}
\label{fig2}
\end{figure}

For sufficiently large $r$ (strictly speaking, for sufficiently large 
$|\boldom |$),
the magnetic dipole approximation breaks down, and the field~(6) can no
longer be used.  The field lines are twisted about $\bf\Omega$ by the 
rotation near the lightsphere, and the particle 
motion is more complex~\cite{gold69}.
Gravitational attraction is negligible compared to the
magnetic force for such energetic particles, even far from the sphere.

\section*{3. LOCAL SINGLE-BODY STATES}

The local Landau states can be obtained by expanding the energy-momentum 
constraint in a Taylor series around $\bf{r}$ = $\bf{\bar{r}},$ i.e., let 
$\bf{r}$ go
to $\bf{\bar{r}}$ + $\bf{\Delta r}.$  The local point is defined by its
spherical coordinates $(\bar{r},\bar{\theta},\bar{\phi}).$  We retain terms
up to second order in $\bf{\Delta r}$ and momentum ${\bf P},$ but leave the
local metric $g_{\mu\nu}(\bf{\bar{r}})$ constant.

\subsection*{3.1 Local coordinates and field} 

We define a local Cartesian coordinate system so that the $z$ co-ordinate is
along the local magnetic field vector, the $x$ coordinate is orthogonal to
$z$ and pointing outward, and the $y$ coordinate is azimuthal.

The dipole magnetic field is
\begin{eqnarray}
{\bf B} = (B_0/2)\Big(\frac{R}{r}\Big)^3\big[ 2\cos\theta~{\bf\hat{r}}
+ \sin\theta~{\bf\hat{\boldthet}}\big]\quad .
\end{eqnarray}
In terms of the local displacement $\bf{\Delta r},$ the new coordinates are 
given by
\begin{eqnarray}
x &=& \frac{{\Delta r}\sin\theta - 2{r\Delta\theta}\cos\theta}{\sqrt{1 +
3\cos^2\theta}}\quad , \nonumber \\
z &=& \frac{2{\Delta r}\cos\theta + {r\Delta\theta}\sin\theta}{\sqrt{1 +
3\cos^2\theta}}\quad , \nonumber \\
y &=& -r\sin\theta{\Delta\phi}\quad ,
\end{eqnarray}
while the local Cartesian basis is
\begin{eqnarray}
\bf{\hat x} & = & \frac{\sin\theta~{\bf\hat{r}} - 
2\cos\theta~{\bf\hat{\boldthet}}}
{\sqrt{1 + 3\cos^2\theta}}\quad , \nonumber \\
\bf{\hat z} & = & \frac{2\cos\theta~{\bf\hat{r}} + 
\sin\theta~{\bf\hat{\boldthet}}}
{\sqrt{1 + 3\cos^2\theta}}\quad , \nonumber \\
\bf{\hat y} & = & -{\bf\hat{\boldphi}}\quad .
\end{eqnarray}
The bar on ${\bf\bar{r}}$ is now omitted unless needed.

\subsection*{3.2 Local Landau states}

The constraint can be rewritten in the local coordinates as 
\begin{eqnarray}
\lefteqn{\epsilon^2-\Pi^2_x - \Pi^2_z - (\Pi^\prime_y-\beta x/r)^2 -1 +}
\nonumber\\
 & & 2\omega_\phi \epsilon(\Pi^\prime_y-\beta x/r) 
 - 2\omega_\theta \epsilon \frac{\sin\theta~\Pi_z
    -2\cos\theta~\Pi_x}{\sqrt{1 + 3\cos^2\theta}} + 
\nonumber\\
 & & 2\omega_\phi \epsilon\Big(\frac{\beta \cos\theta }
{r^2\sin\theta
 (1+3\cos^2\theta)^{3/2}}\Big)[ 4\cos\theta~x^2 - 
\nonumber\\
 & &3\cos\theta\sin^2\theta~z^2
 + 2\sin\theta (3\cos^2\theta - 1 )~xz ] + 
\nonumber\\
 & & (\Pi^\prime_y-\beta x/r)^2\omega^2_\phi + 
\nonumber\\
 & & \frac{(\sin\theta~\Pi_z - 2\cos\theta~\Pi_x)^2}{1+3\cos^2\theta}
\omega^2_\theta = 0\quad .
\end{eqnarray}
Here we use dimensionless local magnetic field strength, energy, and momenta 
given by
\begin{eqnarray}
\beta \equiv \frac{|q{\bf B}|r}{2mc^2} & = &\beta_0\Big(\frac{R}{r}\Big)^2
\sqrt{1+3\cos^2\theta}\quad , \nonumber \\
\epsilon & = & E/mc^2 \quad, \nonumber \\
\Pi_{x,z} & = & P_{x,z}/mc \quad.
\end{eqnarray}
$\Pi^\prime_y $ is the new canonical momentum given by the transformation
\begin{eqnarray}
\Pi^\prime_y = -\frac{P_\phi}{mcr\sin\theta}
+ \frac{qB_0R^3\sin\theta}{2mc^2r^2}\quad . 
\end{eqnarray}
\noindent The second term is a shift which is constant in the local 
coordinate system.
The first line of the constraint~(20) gives the classical version of the
Landau system in terms of the local magnetic field ${\bf B}({\bf\bar{r}}),$ 
which also defines the local $z$ axis.  The other terms are linear and 
quadratic corrections (Coriolis and centrifugal terms) due to the local 
$\boldom$ components $\omega_\phi$ and $\omega_\theta$.

Neglecting rotational effects yields the classical cyclotron 
motion~\cite{land75}.
In the semiclassical form, these are the Landau 
orbitals, with energy eigenvalues
\begin{eqnarray}
\epsilon^2_0 = 1 + \Pi^2_z + \Big(\frac{\hbar}{mc}\Big)
\Big(\frac{2\beta}{r}\Big)[ 2n_L + 1 - {\rm sgn}(q)\sigma ]\quad ,
\end{eqnarray}
where the principal Landau quantum number is $n_L$ = 0, 1, 2, ... 
and the spin
$\sigma$ has been included, $\sigma$ = 
$\pm 1$~\cite{land77,bere82}. For
${\rm sgn}(q)<$ 0, the state $(n_L,\sigma = 1)$ is degenerate with $(n_L+1,
\sigma = -1).$  The 
Landau orbitals are characterized by length and momentum scales $L$ = 
$\sqrt{\hbar c/(2|q{\bf B}|)}$ and $P_L$ = $\sqrt{\hbar|q{\bf B}|/(2c)}$
in the $(x,y)$ plane transverse to the field.  The longitudinal momentum $P_z$
= $mc\Pi_z$ is an eigenvalue continuous over the range $-\infty <$ $P_z$ 
$< +\infty$, as the motion in that direction is force-free.  Note that the 
magnetic factor proportional to $\beta$ is suppressed by the ratio of the 
Compton wavelength $\hbar /(mc)$ to the spherical distance $r.$  The effect of
the magnetic field in the energy is thus negligible compared to $mc^2$ unless 
$\beta$ is very large, as seen in paper~I.  Single-body Landau states of 
fermions are unstable against radiation, unless the lower energy states are 
already filled, as they are in Sect.~IV below.
  
We now consider the effect of rotation as a first-order perturbation in
$\omegabar ,$ approaching the problem semiclassically.  (The terms quadratic
in $\omegabar$ are much smaller.)  The unperturbed orbitals are Landau states
in
the $(x,y)$ transverse plane.  We keep the terms linear in $\boldom$ (Coriolis
effect) and average them over one Landau orbit, as in paper~I, Sect.~V.  
But unlike that case,
this averaging is over a {\it local} microscopic orbit, not a macroscopic
orbit over the whole sphere.  The terms linear in $\omegabar$ thus do not
average to zero in general.  The modified classical energy-momentum constraint 
is then
\begin{eqnarray}
\lefteqn{\epsilon^2-\Pi^2_x - \Pi^2_z - (\Pi^\prime_y-\beta x/r)^2 -1 -} 
\nonumber \\
 & & 2\omega_\theta \epsilon 
\frac{\sin\theta~\Pi_z}{\sqrt{1 + 3\cos^2\theta}} + 
\nonumber \\
 & & 2\omega_\phi\epsilon\Big(\frac{\beta\cos\theta}
{r^2\sin\theta
 (1+3\cos^2\theta)^{3/2}}\Big)\times
\nonumber\\
 & & [4\cos\theta~x^2 - 3\cos\theta\sin^2\theta~z^2] 
= 0\quad ,
\end{eqnarray}
noting that $\langle x\rangle_0$ = $\langle y\rangle_0$ = 
$\langle\Pi_x\rangle_0$ = $\langle\Pi_y\rangle_0$ = 0, averaged over an
unperturbed cyclotron orbit.  These orbits have semiclassical radius $a(n_L)$ =
$2L\sqrt{n_L+1/2}.$  The semiclassical result for the Landau energies corrected
through ${\cal O}(\omegabar )$ is then
\begin{eqnarray}
\epsilon^2_1 & = & \epsilon^2_0
 + 2\omega_\theta\epsilon_0\sin\theta~\Pi_z/\sqrt{1+3\cos^2\theta} -
\nonumber \\
 & &  2\omega_\phi\epsilon_0\Big(\frac{\beta\cos\theta}
{r^2\sin\theta
 (1+3\cos^2\theta)^{3/2}}\Big)\times
\nonumber\\
 & & [4(2n_L+1)\cos\theta~L^2 - 
3\cos\theta\sin^2\theta~z^2_C]~,
\end{eqnarray}
where we have used $\langle\Pi_z\rangle_0$ = $\Pi_z,$ $\langle x^2\rangle_0$ =
$a^2(n_L)/2,$ and a distance cutoff $z_C$ for motion in the $z$ direction.  
(This cutoff is discussed further in Sect.~IV below.) The last term can be
neglected if the Landau orbits are much smaller than the spherical distance
$r:$ $L,$ $z_C<<$ $r,$ as it is another factor of $L/r$ smaller than the
zeroth-order term.  The energies depend on $\Pi^2_z$ at zeroth order but
receive a correction proportional to $\Pi_z$ at first order in $\omegabar ,$
breaking the symmetry $\Pi_z$ $\rightarrow$ $-\Pi_z.$  The Landau pole states
of paper~I can be recovered if $\sin\theta\rightarrow$ 0, $r$ = $R,$
identifying $n_L$ with the old $n_\theta$ and setting $\Pi_z$ = 0.

The error arising from neglecting the cubic terms in the Taylor expansion can 
be estimated and varies as $\sim |{\bf{\Delta r}}|/r$ times the quadratic 
terms.
That is, the cubic and higher terms in the expansion are suppressed by
additional powers of the Landau length $L$ over the sphere size $r.$

\section*{4. LOCAL MANY-BODY STATES}

\subsection*{4.1 Density of states}

The full density of states is a product of four factors: the density of Landau
states, the degeneracy factor ${\cal D}_\perp$ in the $(x,y)$ transverse 
plane for each Landau state, the density of longitudinal states 
${\cal D}_\parallel$ for $z$ motion, and the spin factor (one for the ground 
state $n_L$ = 0, two otherwise).

The degeneracy of a given Landau state $n_L$ in the transverse plane is
\begin{eqnarray}
{\cal D}_\perp = \frac{|q{\bf B}|}{2\pi\hbar c}
\end{eqnarray}
per unit planar area, a result valid in both 
non-relativistic~\cite{ibac95} and
relativistic regimes, while the longitudinal motion contributes a factor
\begin{eqnarray}
{\cal D}_\parallel = \frac{mc~d\Pi_z}{2\pi\hbar}
\end{eqnarray}
per unit longitudinal length.  Thus the number of states, including the spin 
factor, is
\begin{eqnarray}
\frac{d^2{\cal N}}{d{\cal S}~dz} = (2 - \delta_{n_L,0})\cdot{\cal D}_\perp
\cdot{\cal D}_\parallel\quad ,
\end{eqnarray}
per unit transverse surface area $d{\cal S}$ and unit longitudinal length 
$dz.$

In the semiclassical limit, where $n_L$ is quasi-continuous, the
density of Landau states per unit energy is given in dimensionless form by
\begin{eqnarray}
\lefteqn{\frac{dn_L}{d\epsilon} = \Big(\frac{mcr}{4\beta\hbar}\Big)
[1+(2\omega_\theta/\sqrt{1+3\cos^2\theta})]^{-1}\times}
\nonumber\\
 & & [1 + \Pi^2_z + 2\hbar\beta/(mcr)[2n_L + 1 - {\rm sgn}(q)\sigma ] +
\nonumber\\
 & & 2\omega_\theta\epsilon_0\sin\theta~[\Pi_z/\sqrt{1+3\cos^2\theta}]\quad .
\end{eqnarray}

\subsection*{4.2 Thomas-Fermi approximation}

The Thomas-Fermi method approximates quantum many-body fermion states in a 
varying
potential with local states defined by a 
locally constant field~\cite{land77}.
In our case, the local electron states
are filled up to some highest and partially-filled Landau
level $n^*_L$ by the charge carriers, assumed here to be electrons.
That part $\zeta^2$ of the squared energy $\epsilon^2$ arising from the 
$\Pi_z$ terms alone,
\begin{eqnarray}
\zeta^2(\Pi_z) = \Pi^2_z + 2\omega_\theta\epsilon\sin\theta~\Pi_z/\sqrt{
1+3\cos^2\theta}\quad ,\nonumber
\end{eqnarray}
must be cut off at some maximum, as must the corresponding $z$ motion.
In a real system, the cutoffs are provided naturally by the presence of lattice
ions: $|z_C|$ $\ga$ $\hbar /(Z_{\rm eff}\alpha mc)$ (Bohr length) and 
$|\Pi_z|$ $\ga$ $Z_{\rm eff}\alpha ,$ where $Z_{\rm eff}$ is an effective
(screened) positive ionic charge~\cite{land77}. In the realistic 
case the 
$z_C$ term is unnecessary; assume that $\zeta^2 <$ $\zeta^2_F$ = 
$(Z_{\rm eff}\alpha )^2.$  Then the longitudinal momentum $\Pi_z$
is cut off asymmetrically at $\Pi_z$ = $\Pi^+_z >$ 0 and $\Pi^-_z <$ 0, with
$\Pi^+_z\neq$ $-\Pi^-_z$ if $\omegabar\neq$ 0 (Fig.~3).

\begin{figure} 
\vspace{11pc}
\caption{
Momentum-dependent part $\zeta^2$ of the squared energy $\epsilon^2$ as a 
function of
longitudinal momentum $\Pi_z,$ including ${\cal O}(\omegabar )$ correction,
showing the solutions $\Pi^{\pm}_z$ of $\zeta^2$ = $\zeta^2_F.$}
\label{fig3}
\end{figure}

The electron number density $n_e$ at any point is given by
\begin{eqnarray}
n_e=\frac{|q{\bf B}|}{2\pi\hbar c}[1+2(n_L^*-1) + 2\nu]\cdot\Delta \Pi_z 
(mc/2\pi\hbar)
\quad ,
\end{eqnarray}
where $\nu$ is the partial filling factor of the highest Landau level $n^*_L,$
$0\le$ $\nu$ $\le 1,$ and $n^*_L$ is assumed $\ge $ 1.  (If $n^*_L$ = 0,
the entire term in brackets is replaced simply by $\nu .$)
Because of the $z$ degree of freedom, each Landau level is actually a band,
with lowest energy level at some $\Pi_z\neq$ 0, whose sign depends on that
of $\omega_\theta .$  In addition, there is the planar degeneracy, modified
by the partial filling factor $\nu$ in the highest Landau level.  If $n_e$ is
specified along with $\Delta\Pi_z,$ then $n^*_L$, $\nu,$ and $\epsilon_F$
can be determined.  Procedural details are found in the Appendix.

At zero temperature (the case examined in this paper), the electrons cannot
radiate into already-filled lower-energy single-body states.  Radiation occurs
only if fermions are externally excited above the Fermi energy.

If $n_e$ is held constant over a sphere of radius $r,$ then $n^*_L,$ $\nu ,$
and $\epsilon_F$ change over that surface as $|{\bf B}|$ changes.
The variation of $n_L^*$ and $\nu$ over the surface is given by
\begin{eqnarray}
n_L^* & = &  {\rm Int}\Bigl[\big(\frac{2\pi\hbar}{mc}\big)^3\big(\frac{n_e}
{\Delta \Pi_z}\big)
\big(\frac{B_c}{B_0\sqrt{\alpha}}\big)~\big(\frac{r}{R}\big)^3\times
\nonumber\\
 & &\frac{1}{4\sqrt{1+3\cos^2\theta}}+\frac{1}{2} \Bigr]\quad, 
\nonumber\\
\nu & = & \big(\frac{2\pi\hbar}{mc}\big)^3\big(\frac{n_e}{\Delta \Pi_z}\big)
\big(\frac{B_c}{B_0\sqrt{\alpha}}\big)\times
\nonumber\\
 & &\frac{1}{4\sqrt{1+3\cos^2\theta}}+\frac{1}{2} - n_L^*\quad ,
\end{eqnarray}
where ${\rm Int}[f]$ in~(31) denotes the integer value of the function
$f.$  Typical variations of $n^*_L$ and $\nu$ over the sphere are shown in 
Figs.~4 and~5.  The Fermi energy at any point on the surface is
\begin{eqnarray}
\epsilon_F^2 & = 1 + \zeta^2_F + \Big(\frac{B_0\sqrt{\alpha}}{B_c \pi}
\Big)\Big(\frac{R}{r}\Big)^3\big( 2 n_L^* +1 + \sigma\big)\times
\nonumber\\
 & \sqrt{1 +3\cos^2 \theta}\quad ,
\end{eqnarray}
where ${\rm sgn}(q)$ = $-1,$ and $\sigma_F$ = $-1$ for the $n^*_L$ level; the
critical field strength is $B_c$ = $(m^2_e/\pi )\sqrt{c^5/\hbar^3}$ = $1.3
\times 10^{12}$ G, defined in paper~I.  Matching variations of $\epsilon_F$ are
shown in Fig.~6.
As the Fermi energy $\epsilon_F$ varies over a sphere, electron currents flow
as the electrons seek the lowest energy. There are also radial currents (see
Sect.~IV.C below).  

Quantitative details of the subsequent evolution~\cite{rude86} are beyond 
the scope of this
paper, but certain qualitative features are clear.  The 
magnitude and evolution of the currents depend on the electrical conductivity, 
which in turn depends on the nuclear crystal and the magnetic field.
(Relativistic electrons travel at essentially the speed of light, leading to 
saturated current densities of ${\cal O}(n_ec),$ apart from magnetic
effects on the density of states~\cite{gelb98}). As the 
magnetic field and
rotation affect the electron Fermi energy, the beta equilibria of 
neutrons~\cite{canu75} 
and muons are also affected above their respective thresholds.
This last effect cannot be included without 
revising the hadronic equation of 
state~\cite{chak97}. The Fermi energy will be
unequal over the sphere at first, a limit expected to apply to young neutron
star crusts.  With time, $\epsilon_F$ will equilibrate
to the same value everywhere, and the ${\bf B}$ and $n_e,$ $n_p$ profiles will
change in parallel.

\begin{figure} 
\vspace{33pc}
\caption{
The last, partially-filled Landau level $n^*_L$ on a spherical surface,
at constant density, defined by~Eq.~(31); $B_0$ = 10$B_c.$
(a) $\rho$ = 10 g/cm$^3;$ (b) $\rho$ = 10$^7$ g/cm$^3;$ (c) $\rho$ = 
10$^{13}$ g/cm$^3.$ $\theta$ and $\phi$ in radians.}
\label{fig4}
\end{figure}

\begin{figure} 
\vspace{33pc}
\caption{
The partial filling factor $\nu$ of the last Landau level $n^*_L$ on a spherical
surface at constant density, defined by~Eq.~(31); $B_0$ = 10$B_c.$
(a) $\rho$ = 10 g/cm$^3;$ (b) $\rho$ = 10$^7$ g/cm$^3;$ (c) $\rho$ = 
10$^{13}$ g/cm$^3.$  $\theta$ and $\phi$ in radians.}
\label{fig5}
\end{figure}

\begin{figure} 
\vspace{33pc}
\caption{
The dimensionless Fermi energy $\epsilon_F$ on spherical a surface at 
constant density, defined by~Eq.~(32); $B_0$ = 10$B_c.$
(a) $\rho$ = 10 g/cm$^3;$ (b) $\rho$ = 10$^7$ g/cm$^3;$ (c) $\rho$ = 
10$^{13}$ g/cm$^3.$  $\theta$ and $\phi$ in radians.}
\label{fig6}
\end{figure}

The opposite limit is the case where the Fermi energy is constant in space
and no currents flow, a situation probably holding for crusts at later times.
Since the field $|{\bf B}|$ varies, the electron
density $n_e$, and thus the positive ion density $n_p$, must also vary.  
This is
the case where the magnetic field is strong enough to dominate the mechanical
structure of dense matter.  We show three
examples of how $n_e$ is redistributed over a sphere of constant $r$ for fixed
$\epsilon_F$ (Fig.~7).  The proton density is $n_p$ = $(Z/Z_{\rm eff})n_e,$ 
and $\rho$ can be inferred from Eq.~(A2).

\begin{figure} 
\vspace{33pc}
\caption{
The electron number density $n_e$ on spherical surface at constant 
Fermi energy, defined by~Eq.~(30); $B_0$ = $B_c.$  Units are inverse
Compton volume.  (a) $\epsilon_F$ = 1.06; (b) $\epsilon_F$ = 5.0;  
(c) $\epsilon_F$ = 1800.  $\theta$ and $\phi$ in radians.}
\label{fig7}
\end{figure}

\subsection*{4.3 Radial structure}

For a simple neutron star, we can assume a given profile of positive ions, and 
thus electrons, with radius $r.$  With the field $|{\bf B}({\bf r})|$ in 
addition, the radial dependence of $\epsilon_F$ can be found.

In Fig.~8 the radial profile is shown for $\epsilon_F$ from $r/R$ = 0.7
to 1.0, with an expanded subfigure of $r/R$ = 0.9998--1.0.  Even at this
detail, the thin surface non-relativistic layer cannot be seen, but the 
quantum regime of discretized $\epsilon_F$ steps is clearly visible.
Fig.~9 shows $\epsilon_F$ as a function of $r/R$ and
$\theta .$  The rotational correction drops out in certain cases.  If
$\sin\phi$ = 0 $({\bf{\bar r}}$ in the ${\bf M}$--${\bf\Omega}$ plane) or 
$\sin\theta_0$ = 0 (no tilt), $\omega_\theta$ vanishes.  
Also, if $\sin\theta$
= 0 $({\bf{\bar r}}$ along the dipole axis), $P_\theta$ $\sim r\sin\theta\cdot 
P_z$ $\rightarrow$ 0, and the Coriolis effect disappears.  Otherwise
the $\epsilon_F$ radial profile is affected dramatically by the rotational
correction for relativistic $\epsilon_F.$

\begin{figure} 
\vspace{22pc}
\caption{
Dimensionless Fermi energy $\epsilon_F$ as function of $r/R$ for
simplified neutron star crust model of text; $\sin\theta$ = $\sin\phi$ = 1,
and $B_0$ = 10$B_c.$  (a) $r/R$ = 0.7--1.00; (b) $r/R$ = 0.9998--1.00.
Terraced steps indicate quantum regime at the surface.}
\label{fig8}
\end{figure}

\begin{figure} 
\vspace{22pc}
\caption{
Dimensionless Fermi energy $\epsilon_F$ as a function of $r/R$ and
$\theta$ for simplified neutron star crust model of text.  (a) $\sin\phi$ = 1,
$r/R$ = 0.7--1.00, $B_0$ = 10$B_c,$ $\sin\theta >$ 0; (b) detail
showing $\sin\theta >$ 0.2.  $\theta$ and $\phi$ in radians.}
\label{fig9}
\end{figure}

For this crust profile, we have used a neutron star model of 
Glendenning~\cite{glen97},
with only electrons, nuclei, and neutrons, and no muons or other hadrons.  The 
radius and mass of the neutron star are $R$ = 11 km and $M$ = 1.55$M_\odot ,$
respectively.  The profile starts at $r/R$ = 0.7, where $\rho$ = $2\times 
10^{13}$ g/cm$^3$ and decreases outward.  As $\epsilon_F$ varies in space,
currents flow in response, a scenario probably relevant to early neutron star
crusts.  (Recall that we ignore the effect of gravity.)  
Again, the neutron and muon beta equilibria are
changed from their zero-field analogues.

Now consider the opposite limit again, probably holding for older crusts, 
a constant $\epsilon_F,$ with $n_e(r)$ determined by a given field profile 
${\bf B}({\bf r}).$  Then $\nu$ must be set independently; here we take 
$\nu$ = 0.1 and 0.9 as illustrative.  For critical field strength,   
the density {\it rises} with $r$, while for even higher fields, the density
is quenched to essentially a constant (Fig.~10).  For the same surface 
value of $\epsilon_F,$ the $n_e$ density of 
Glendenning~\cite{glen97} is too large to 
be shown on the scale of Fig.~10.  Thus, the crust densities are low compared
to the non-magnetized values when the magnetic field is the controlling factor.

\begin{figure} 
\vspace{11pc}
\caption{
The electron number density $n_e$ as a function of $r/R$ for constant 
$\epsilon_F$ = 1.06; $\sin\theta$ = $\sin\phi$ = 1, and $B_0$ = 10$B_c.$  
Solid: $\nu$ = 0.1; dotted: $\nu$ = 0.9.  Density 
profile of simplified neutron star crust model in text is too large to show
on this scale.  Units are inverse Compton volume.}
\label{fig10}
\end{figure}

\section*{5. CONCLUSION}

This concludes the treatment of semiclassical orbitals begun in paper~I.
In this paper, we have found the local one- and many-body states of
relativistic charged particles confined to a sphere with
an intense, rotating magnetic dipole field.

There remain full quantization with the Dirac equation and the inclusion of
the positively-charged lattice structure to determine the local neutron star
matter state.  A full calculation requires a self-consistent treatment of
gravity, nuclear matter, magnetic fields, and currents, with chemical
equilibrium and Coulomb neutrality.  Depending on the degree of lattice
disorder and interelectron forces, various conducting, insulating, or quantum
Hall-like many-body states can arise~\cite{ibac95,frad91}, 
affecting the macroscopic currents inferred in Sect.~IV. 

\section*{ACKNOWLEDGMENTS}

The authors are indebted to Sudheer Maremanda (U.T. Arlington)
for preparing the figures.
We thank Pradeep Kumar (Univ. Florida) for his original 
suggestion and helpful discussions.
This work was supported at the Univ. Florida Institute for 
Fundamental Theory and U.S. 
Department of Energy Contract DE-FG05-86-ER40272;
and by the Research Enhancement Program at U.T. Arlington.

\section*{APPENDIX: MANY-BODY THEORY}

To determine the filling of momentum states labelled by $\Pi_z,$ the cutoff
equation $\zeta^2(\Pi_z)$ = $\zeta^2_F$ must be solved for some given value
of $\zeta^2_F,$ taken here as $(Z_{\rm eff}\alpha )^2.$  At zeroth order
in $\omegabar ,$ this equation is trivial.  Once the rotational corrections
are included, the equation is not only quadratic and linear in $\Pi_z,$ but
has an implicit dependence on $\Pi_z$ through $\epsilon_0.$

In the ${\cal O}(\omegabar )$ correction, our procedure is to take 
$\epsilon^2_0$ = $1 + \epsilon^2_L,$ where $\epsilon^2_L$ is the purely 
two-dimensional Landau term:
\begin{eqnarray}
\epsilon^2_L = \frac{\hbar |q\bf{B}|}{m^2c^3}[2n_L + 1 - {\rm sgn}(q)\sigma ]
\nonumber
\end{eqnarray}
and neglect the $\Pi^2_z$ term, as the
latter is typically much smaller than one.  In that case, the cutoff
equation $\zeta^2(\Pi_z)$ = $\zeta^2_F$ is a simple quadratic with the two
roots $\Pi^{\pm}_z$ and
\begin{eqnarray}
\lefteqn{\Delta\Pi_z = \Pi^+_z - \Pi^-_z =}
\nonumber\\
 & & 2\sqrt{\zeta^2_F + \omega^2_\theta (1 +
\epsilon^2_L)\sin^2\theta /(1+3\cos^2\theta )}\quad
\end{eqnarray}
as the allowed spread of $z$ momenta.  Given a value of $n_e,$ the values
of $\epsilon_F,$ $n^*_L,$ and $\nu$ are found iteratively, starting with
$\omegabar$ = 0, then with this solution used in the ${\cal O}(\omegabar )$
corrections.

In the opposite case, of fixed $\epsilon_F,$ $n^*_L$ is determined, while
we set $\nu$ = 0.1 and 0.9 as illustrative (Fig.~10).  
The density $n_e$ is then found.
As long as $\nu\neq$ 0.5, $n_e$ can fall or rise with $r.$
A complete treatment of the bulk crust requires inclusion of nuclear
matter and gravity~\cite{chak97}, as well
as an interior ${\bf B}$ field profile, not necessarily a dipole.

A semi-realistic spatial profile of electron density requires the proton 
density $n_p$ = $n_e,$ usually determined in terms of mass density $\rho .$  
The ``effective'' electron density $n_e,$ the density available for conduction,
is
\begin{eqnarray}
(2\pi\hbar /mc)^3n_{e, \rm cond} = \frac{Z_{\rm eff}}{A}\cdot 
\frac{1.1\times 10^{-5}\rho}{{\rm g}~{\rm cm}^{-3}}\quad ,
\end{eqnarray}
where $Z_{\rm eff}$ is the number of electrons per nucleus available for
conduction (unbound electrons).  The atomic number $A$ = $N$ + $Z,$
where the neutron number $N$ per nucleus is abnormally large for nuclei in
an electron Fermi sea.  The number density is normalized 
to a Compton volume $(2\pi\hbar /mc)^3.$  From $\rho$ = 10 to about $3\times 
10^4$ g/cm$^3,$ the inner electrons of the atoms remain bound, not 
participating in conduction; in this case, $Z_{\rm eff} <$ $Z$ and can be 
read off from standard atomic structure~\cite{paul70}.  The nuclei are always 
iron $(Z$ = 26 and $Z_{\rm eff} =$ 8, 16, 24) at these 
densities~\cite{glen97,born73}.
For higher densities, the orbitals
of different atoms merge, and $Z_{\rm eff}$ = $Z.$  (We neglect the formation
of partially ionized atoms at ultrahigh densities, assuming that all electrons
are stripped from their nuclei.)  The results do not depend sensitively on
$Z_{\rm eff}.$

For nuclear matter composition at densities below complete nuclear 
dissociation, but for $Z >$ 26, we use the results from 
\cite{bbp71,bps71} (see also \cite{born73}) at
$\rho <$ $2\times 10^{13}$ g/cm$^3.$  Our method can be applied at higher 
densities with free nucleons, but it needs 
to incorporate the presence of muons
and then of heavier strange and non-strange hadrons, and then possibly of
a quark-gluon plasma~\cite{glen97}.
In this paper, only the simplest case of 
electrons and positive ions is examined.

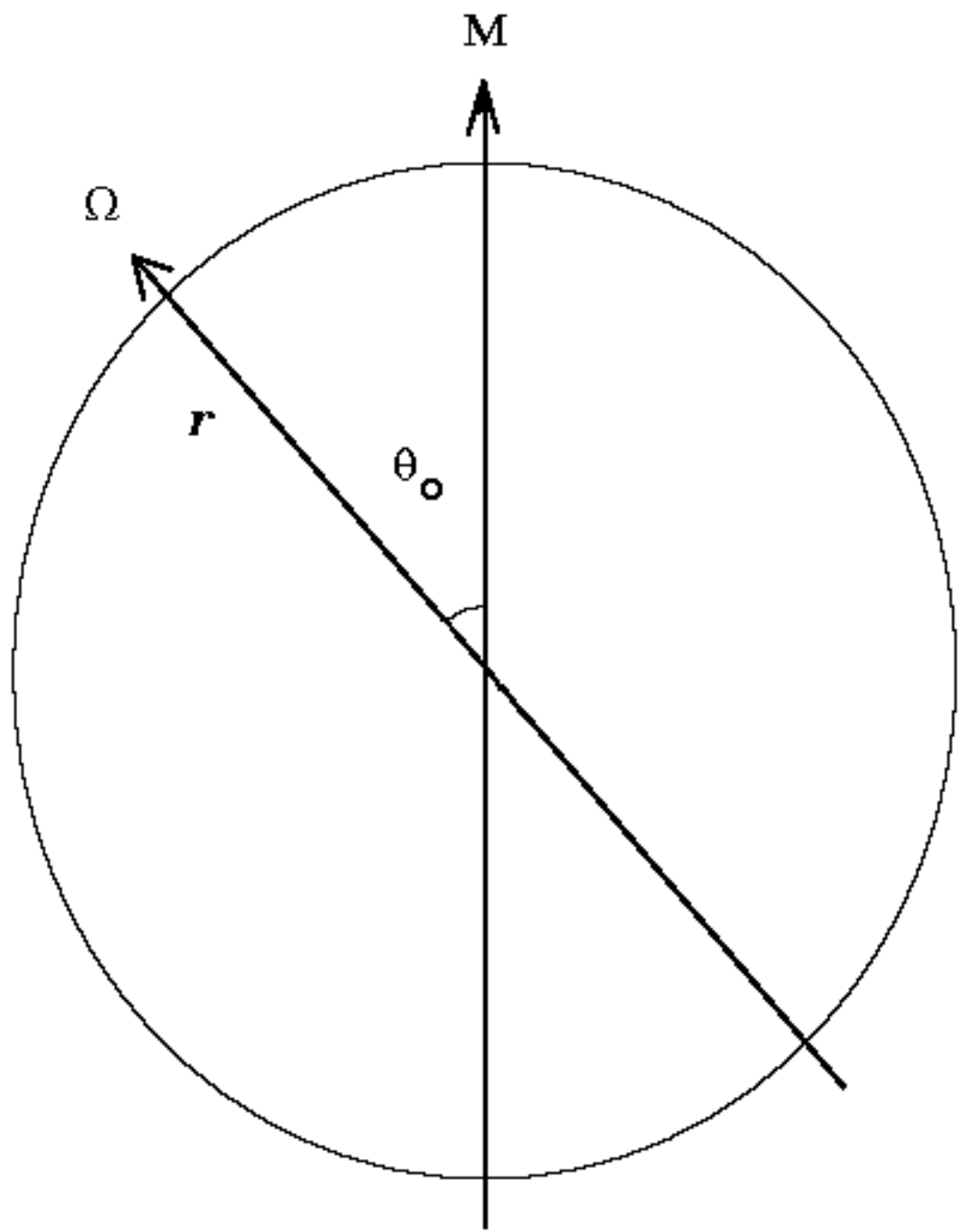
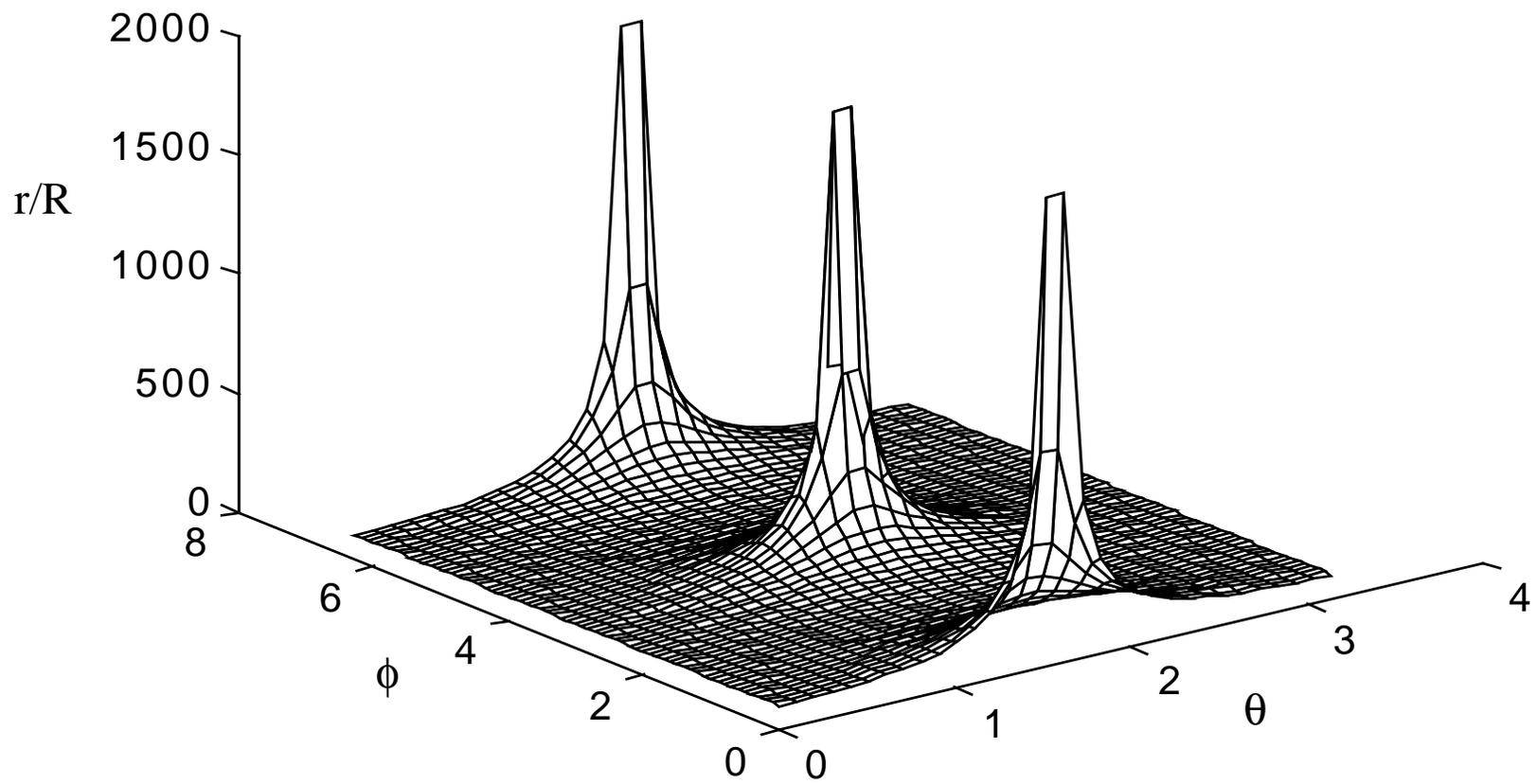
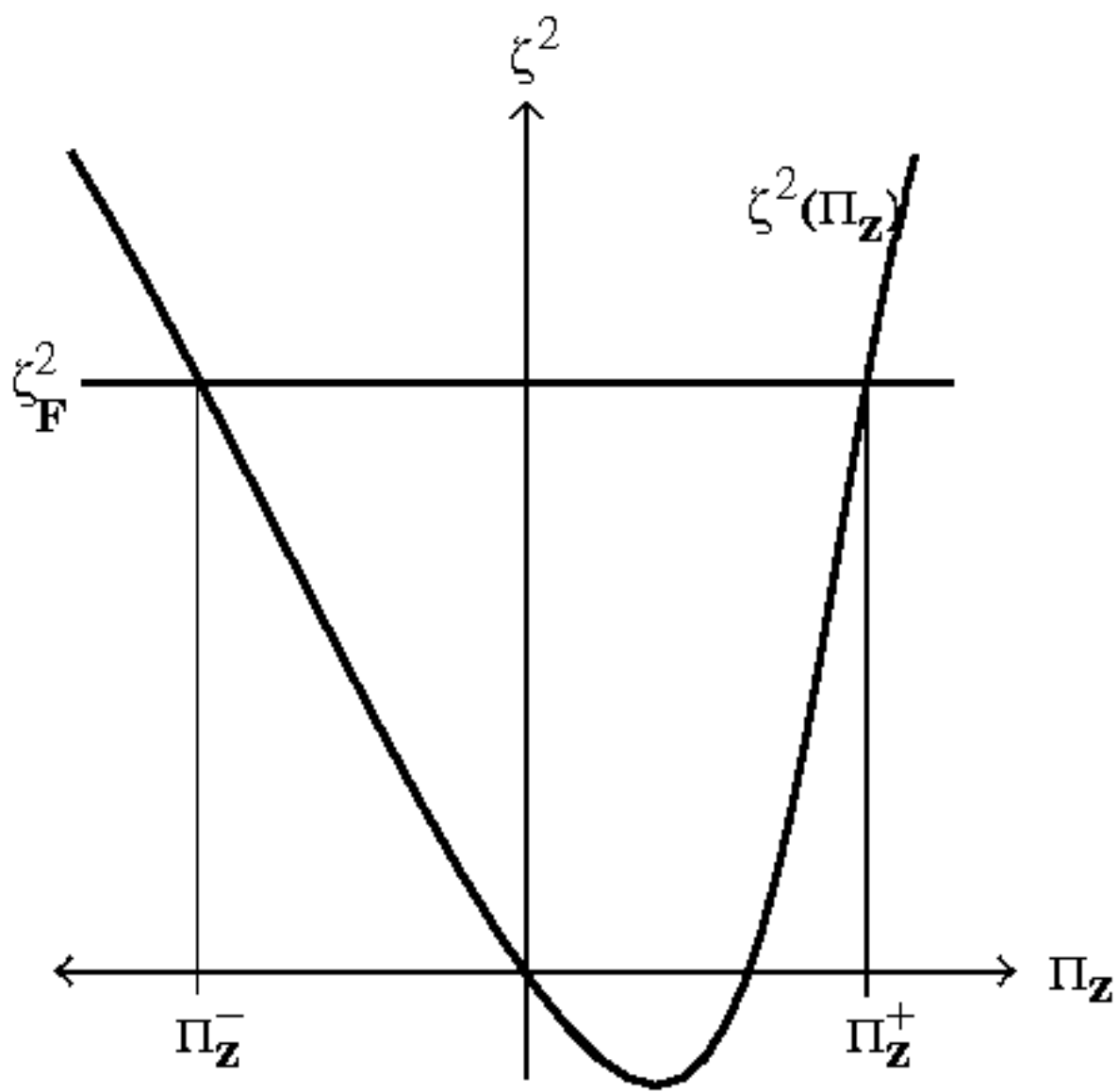
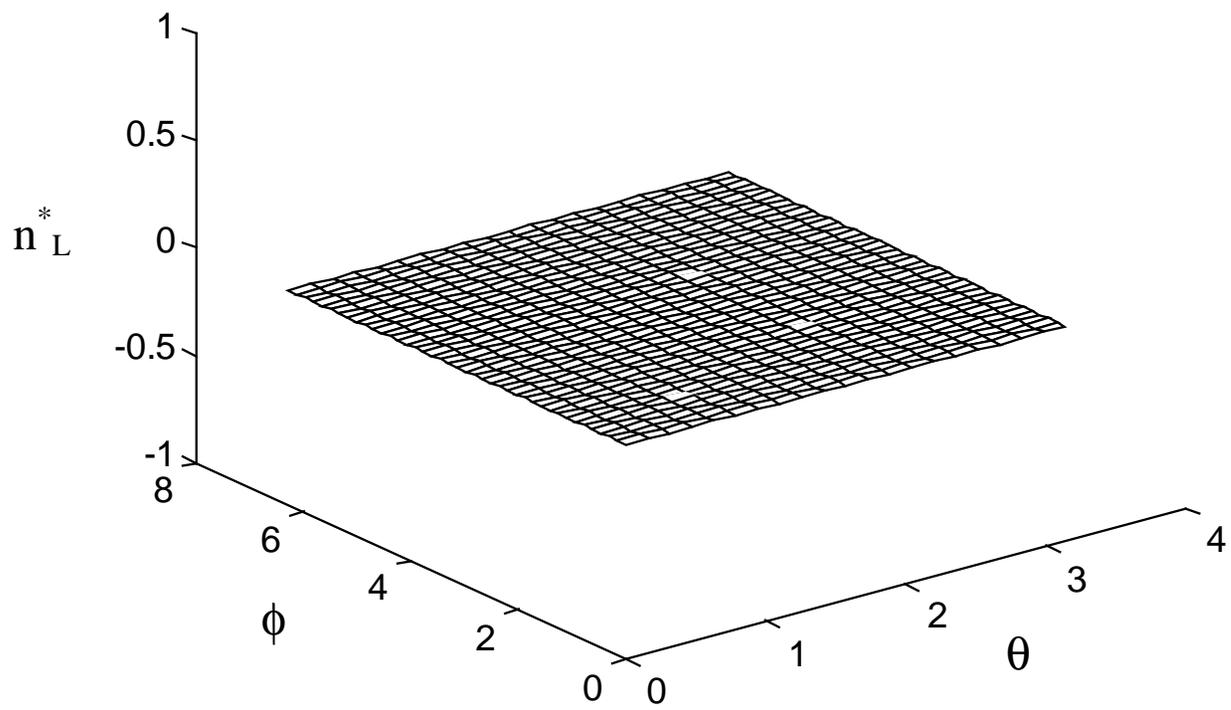
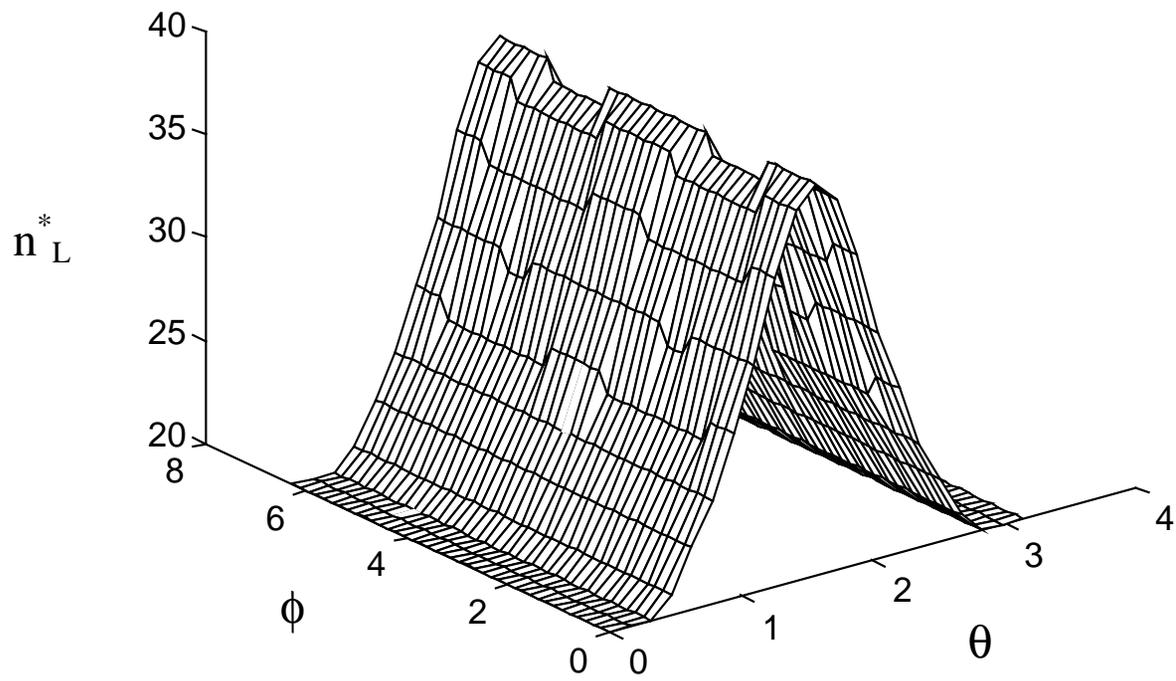
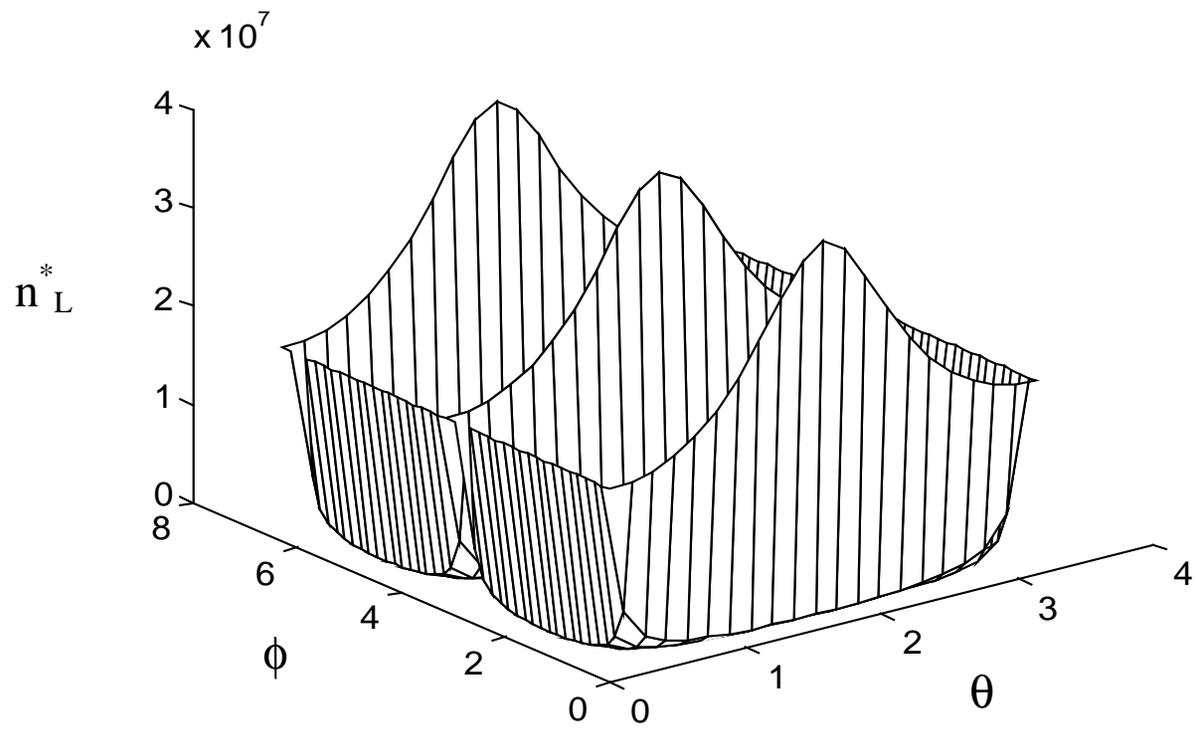
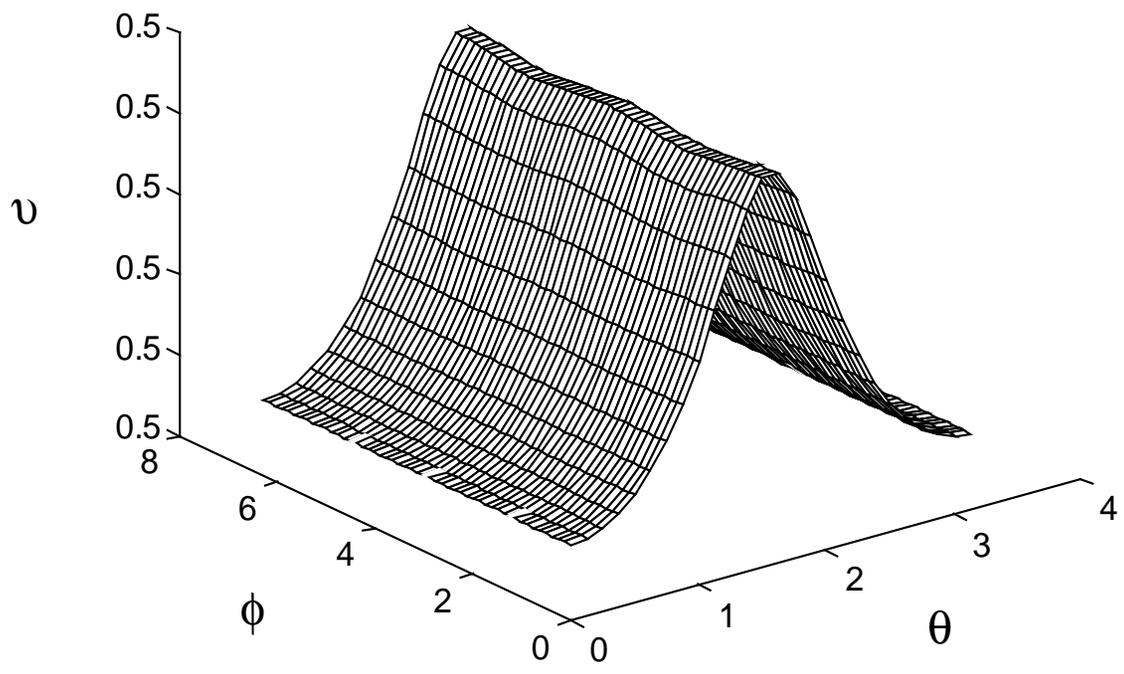
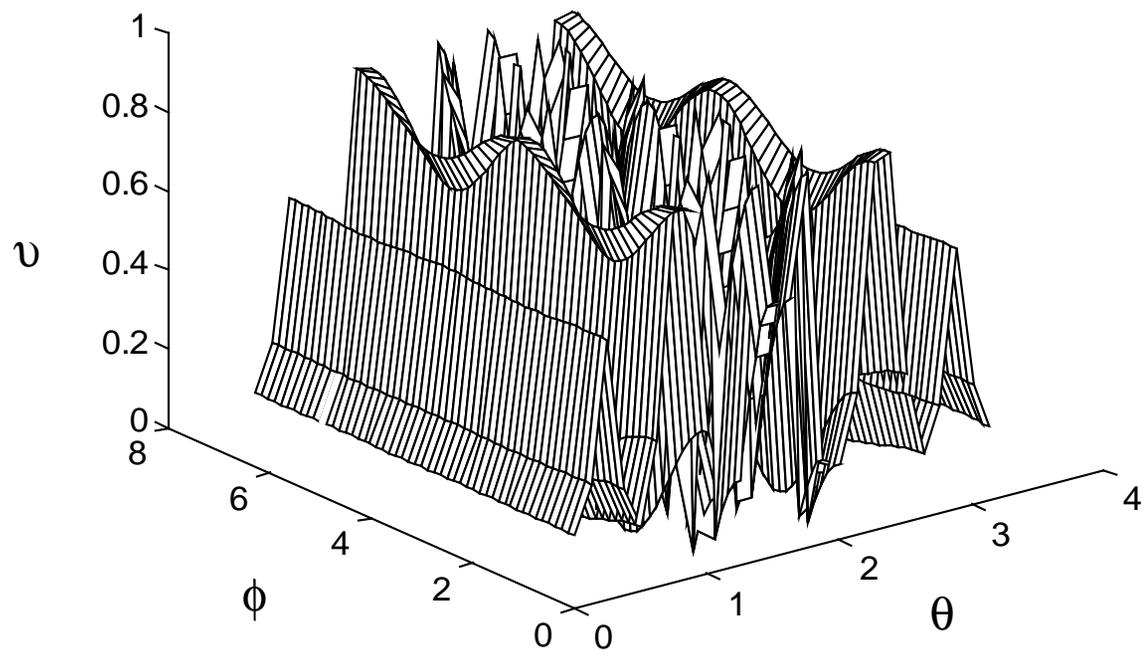
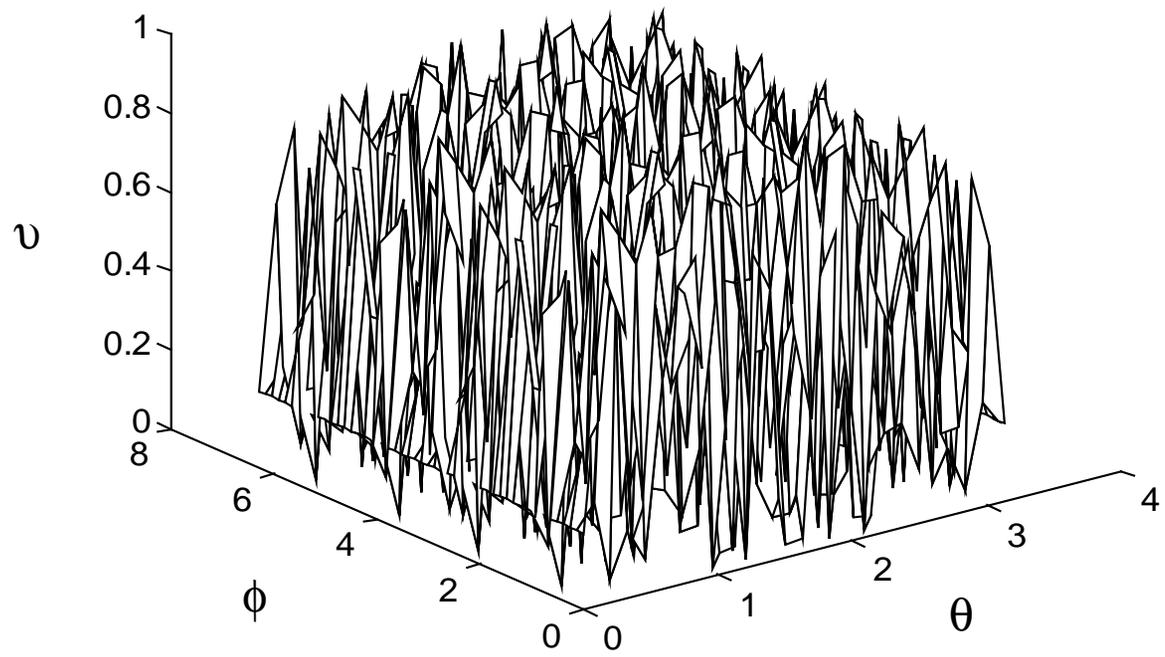
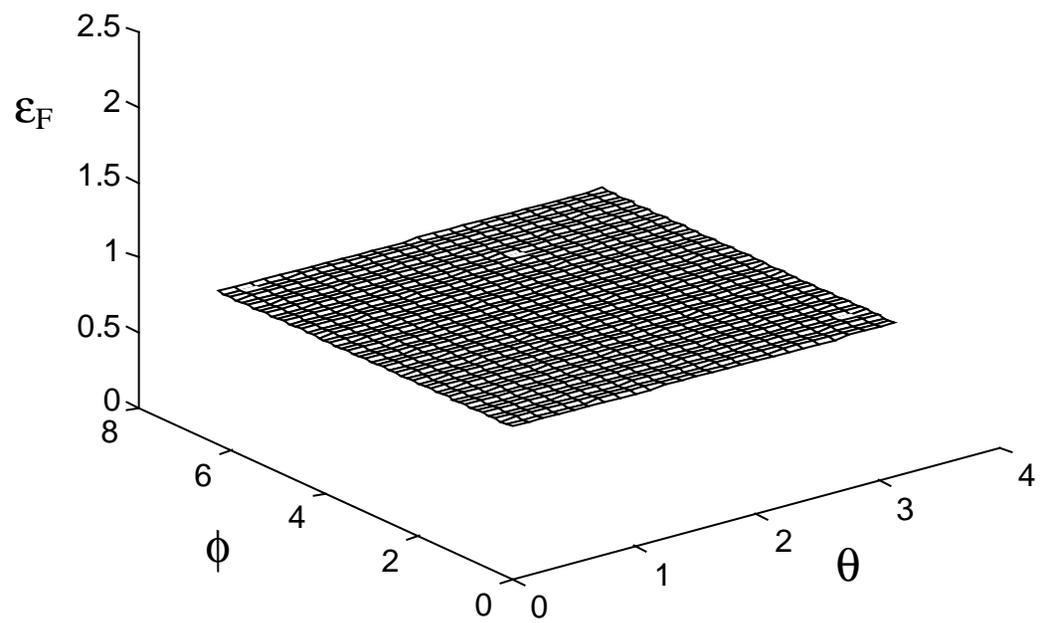
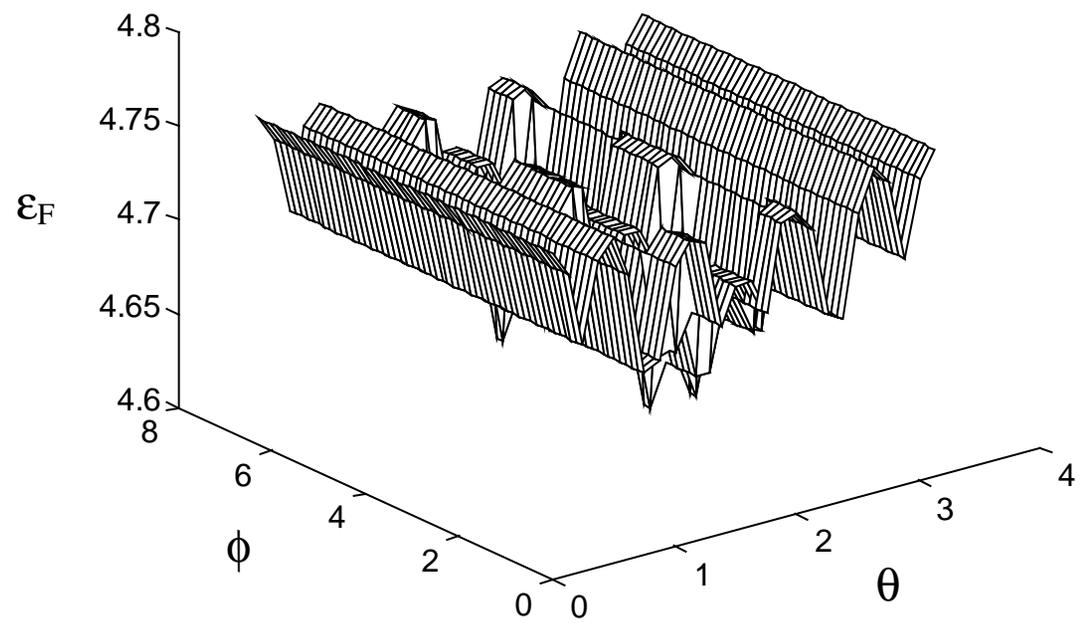
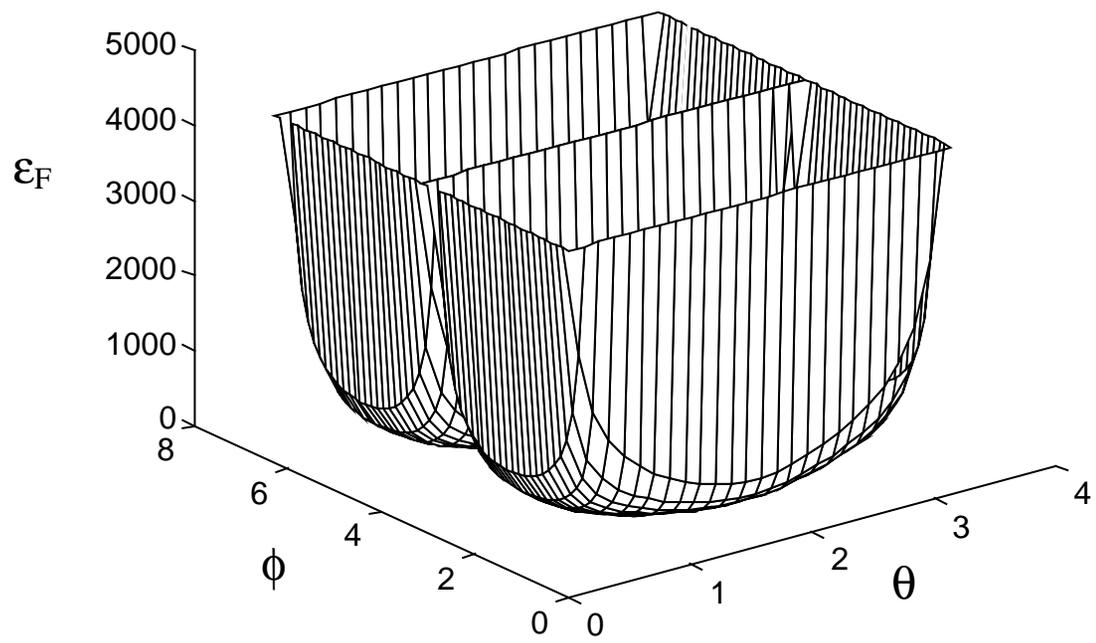
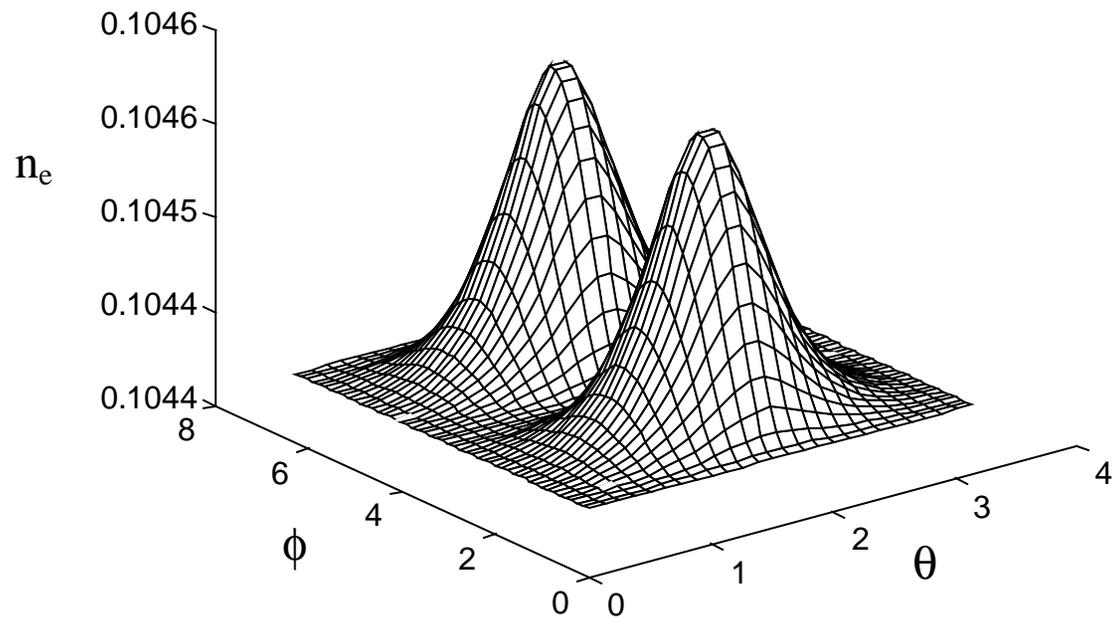
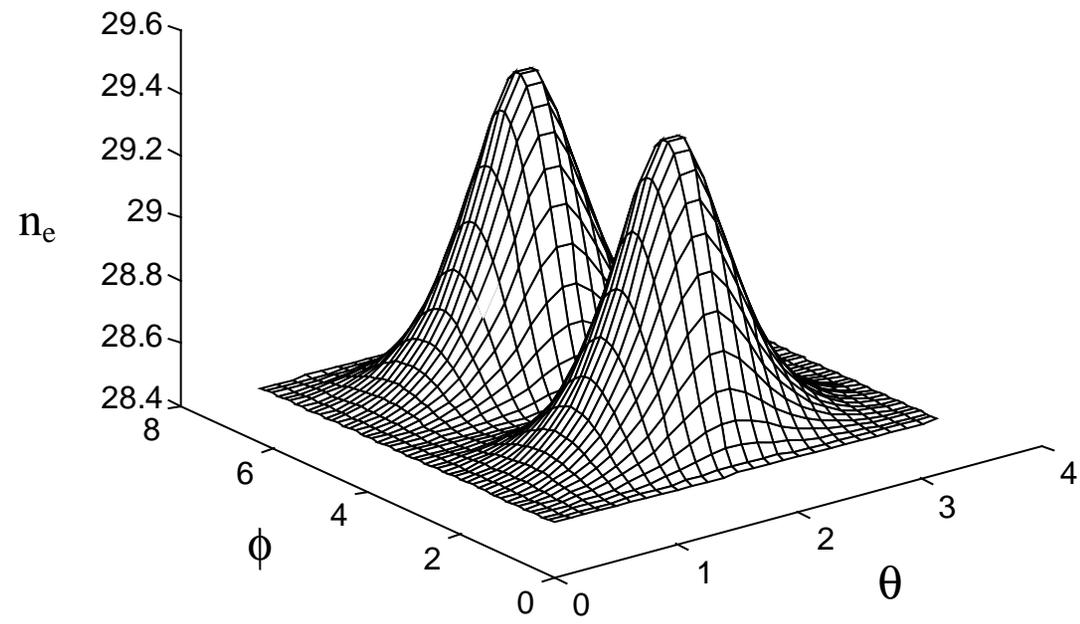
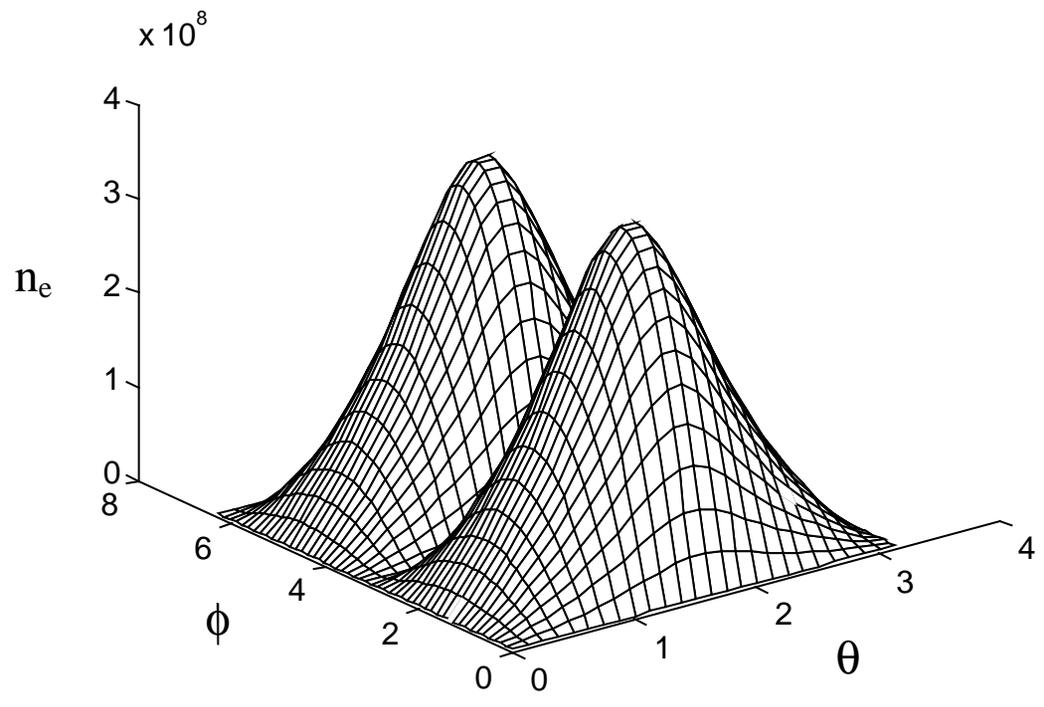
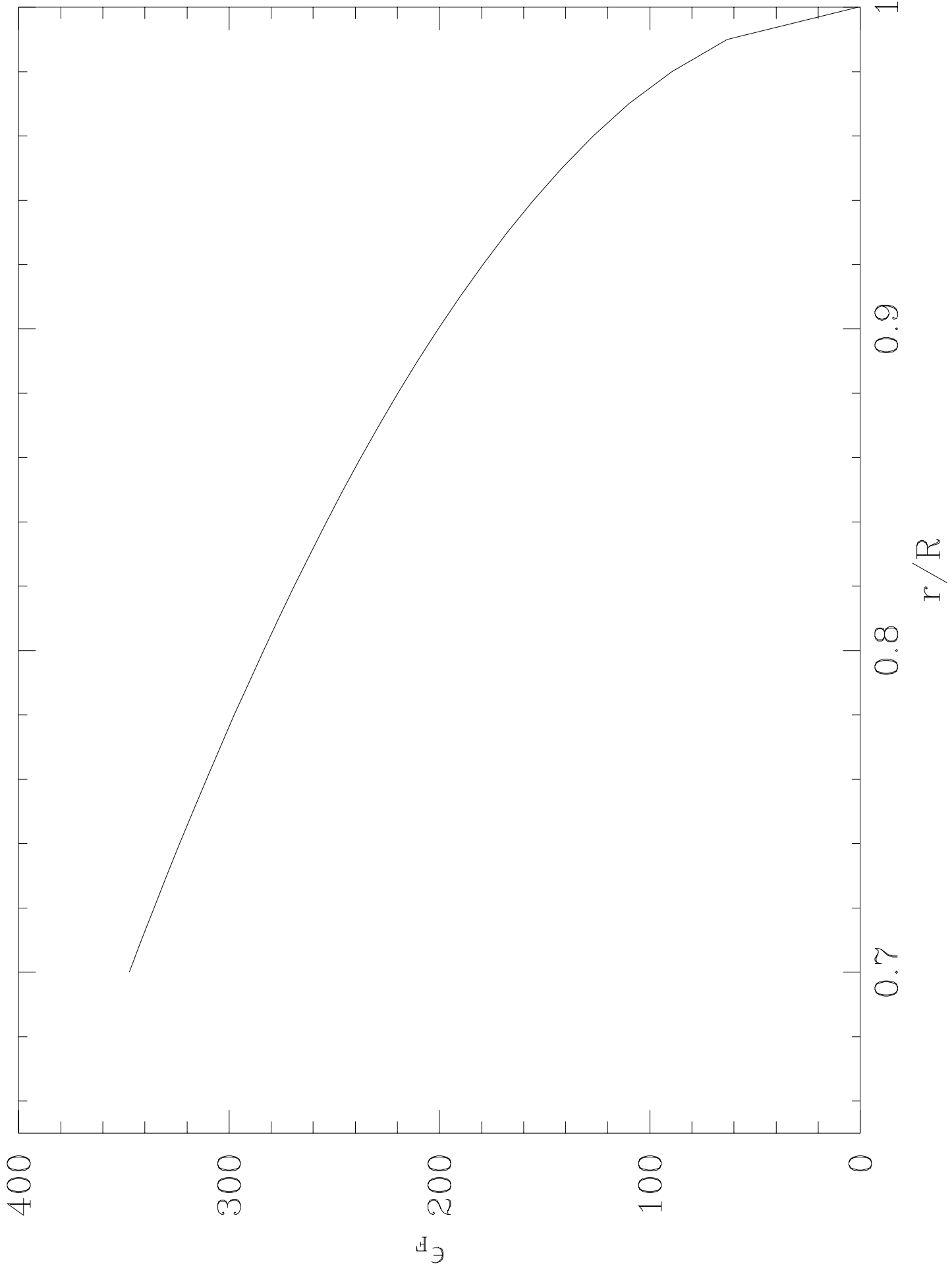
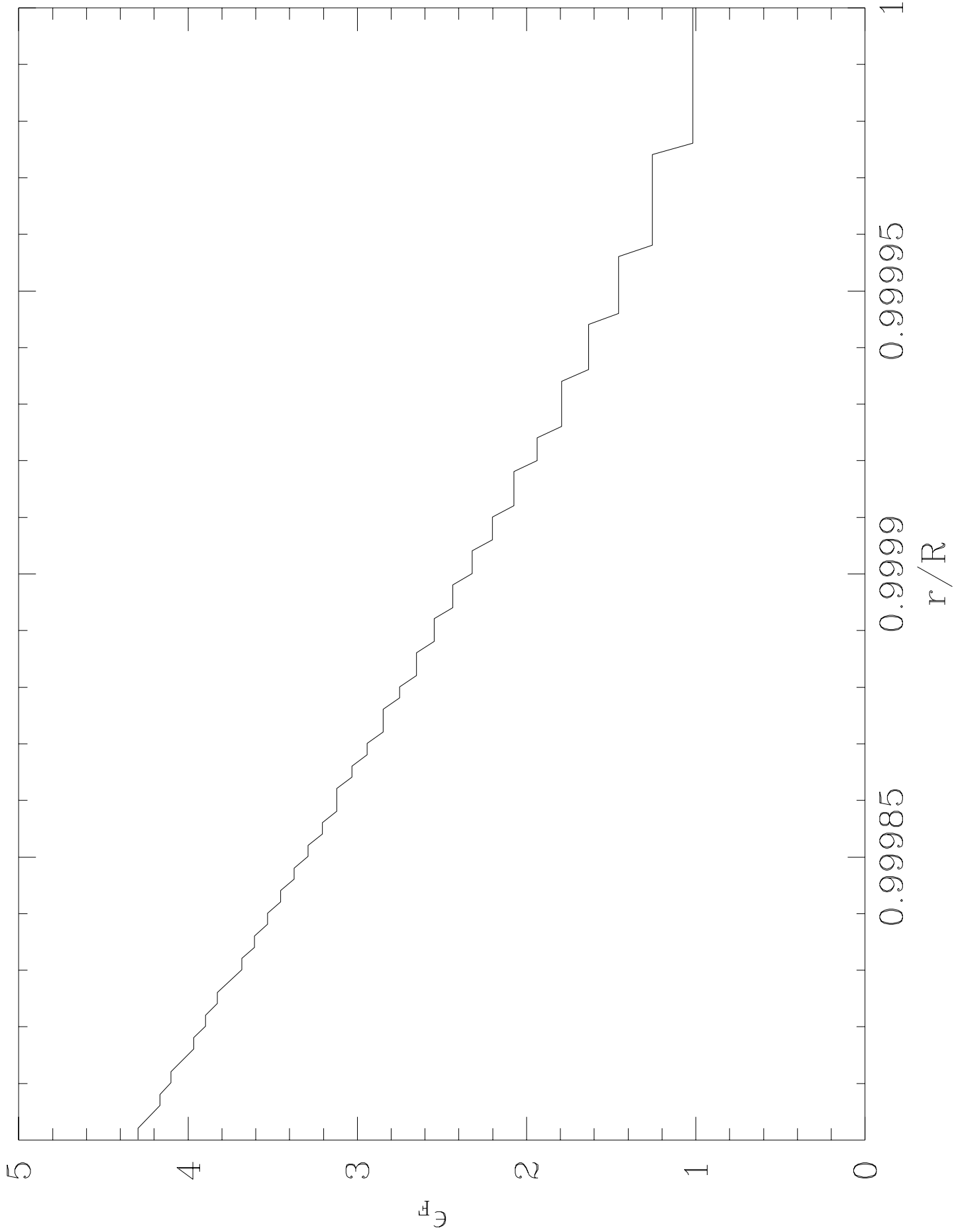
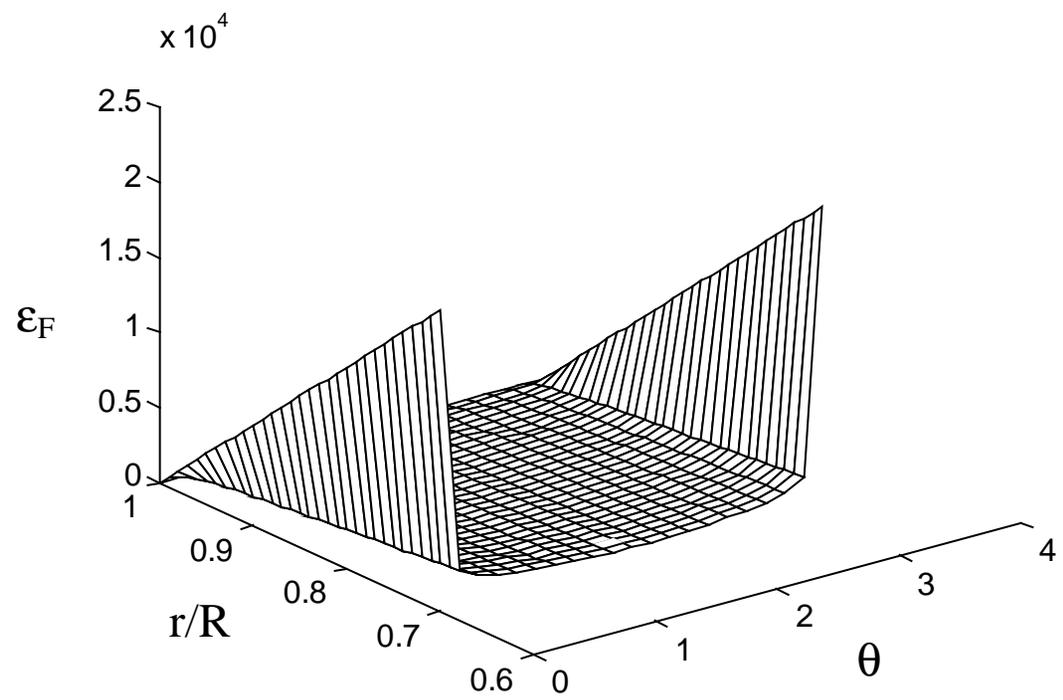
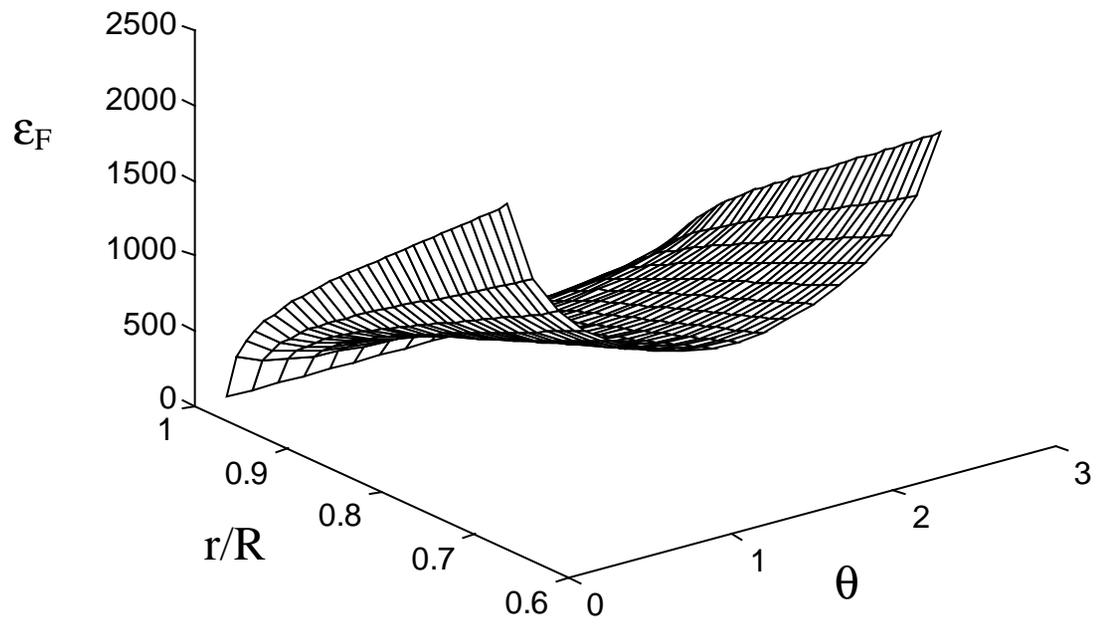
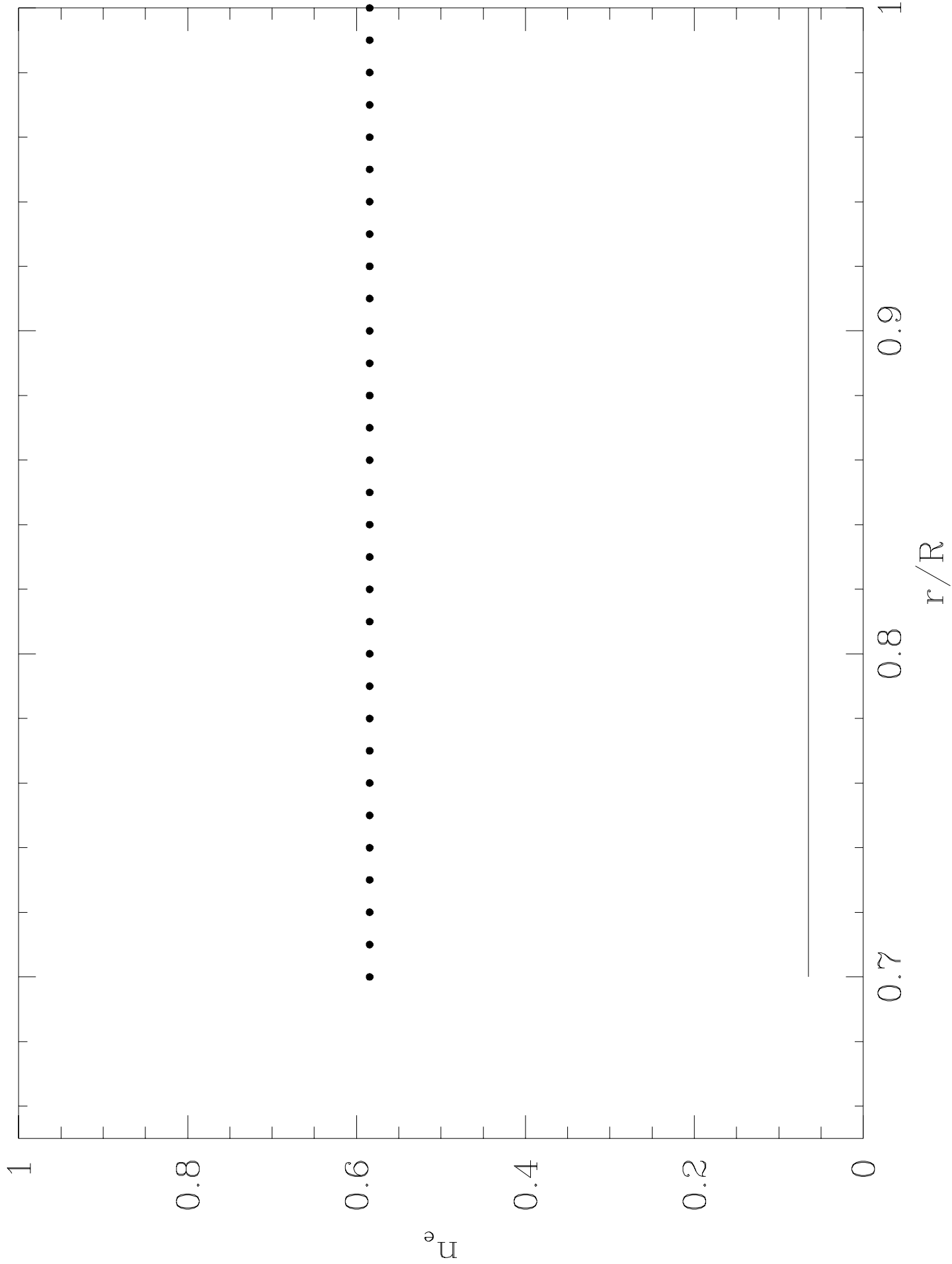
\end{document}